\documentclass[]{spie}  %>>> use for US letter paper
\usepackage{amsmath,amsfonts,amssymb}
\usepackage{graphicx}
\usepackage[colorlinks=true, allcolors=blue]{hyperref}
\usepackage{siunitx}
\newcommand{\SIadj}[2]{\SI[number-unit-product={\text{-}}]{#1}{#2}}
\usepackage{xcolor}

\title{Reflectance measurements of mm-wave absorbers using frequency-domain continuous wave THz spectroscopy}

\author[a]{Gaganpreet Singh}
\author[b]{Rustam Balafendiev}
\author[a,c]{Zeshen Bao}
\author[b]{Thomas J.L.J. Gascard}
\author[a,b]{Jon E. Gudmundsson}
\author[a]{Gagandeep Kaur}
\author[d]{Vid Primožič}

\affil[a]{Oskar Klein Centre, Department of Physics, Stockholm University, AlbaNova, SE-10691 Stockholm, Sweden}
\affil[b]{Science Institute, University of Iceland, 107 Reykjavik, Iceland}
\affil[c]{School of Engineering Sciences, KTH-Royal Institute of Technology, SE-10691 Stockholm, Sweden}
\affil[d]{Faculty of Mathematics and Physics, University of Ljubljana, Ljubljana, Slovenia}

\authorinfo{Further author information: (Send correspondence to G.S. and J.E.G.)\\G.S.: E-mail: gaganpreet.singh@fysik.su.se\\  J.E.G.: E-mail: jegudmunds@hi.is}%, Telephone: +354 525 4626}

\begin{document} 
\maketitle

\begin{abstract}
Due to high dynamic range and ease of use, continuous wave terahertz spectroscopy is an increasingly popular method for optical characterization of components used in cosmic microwave background (CMB) experiments. In this work, we describe an optical testbed that enables simultaneous measurements of transmission and reflection properties of various radiation absorbing dielectric materials, essential components in the reduction of undesired optical loading. To demonstrate the performance of the testbed, we have measured the reflection response of five absorbers commonly used for such applications: TKRAM, carbon- and iron-loaded Stycast, HR10, AN72, and an in-house 3D printed absorber across a frequency range of 100 to 500 GHz, for both S- and P-polarization, with incident angles varying from \ang{15} to \ang{45}. We present results on both the specular and scattered reflection response of these absorbers. 

\end{abstract}


\section{INTRODUCTION}
\label{sec:intro} 

Instruments observing the cosmic microwave background (CMB) are increasingly relying on highly-engineered optical components, including lenses \cite{golec2022simons}, half-wave plates \cite{Komatsu2020}, filters \cite{Tucker2006}, beam-forming elements \cite{Gascard2023}, and absorbers.\cite{singh2019fabrication} Non-idealities in optical systems will cause unwanted reflections and diffractions that can reduce the optical performance of these systems. Absorbers placed in strategic locations within an optical system can help mitigate these effects.\cite{Gudmundsson2021,xu2021simons} Although most of these absorbers are engineered to cover a broad frequency range to match the deployed frequencies for various CMB instruments, the list of publications that describe testing of these absorbers across a wide frequency range and a number of incidence angles is rather limited.\cite{Hemmati1985, Halpern1986, Lamb1997, lonnqvist2006monostatic,Wollack2008} Over the past decade, we have seen an increase in the development of millimeter-wave absorbers with detailed geometries based on both 3D printing and injection molding techniques.\cite{petroff20193d,otsuka2021material,Sun2022} This work will likely continue with the development of the LiteBIRD satellite mission.\cite{LiteBIRD2020} 

A combination of reflection and transmission measurements can be used to constrain the complex constrain optical properties of various dielectrics. In this article, we describe a measurement setup using an ultra-broadband coherent system called Terascan 1550 from TOPTICA Photonics that enables such measurements. We describe the modifications that we have made to the as-purchased system and demonstrate the system performance through measurements of the reflection properties of common absorber materials as function of frequency, angle of incidence, and scattering angle. 

\begin{figure}
    \centering
    \includegraphics[width=0.4\linewidth]{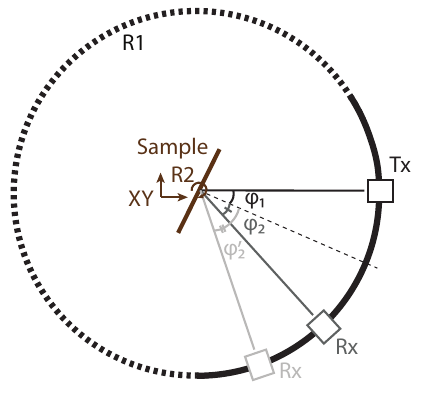}
    \caption{The distance from the center of the parabolic mirrors in the Tx and Rx modules are located \SI{77.5}{\milli\meter} from the sample plane. The sample can be rotated around the R2 axis. The R2 rotation stage is mounted on an XY linear stage with 25 mm of throw. The Rx module is mounted on the R1 stage. The rotation of the R1 stage allows us to measure reflected power from the sample over a range of angles. The solid ring segment indicates roughly the part of the angular region that is used for measurements in this work.
    }
    \label{fig:diagram}
\end{figure}

The Terascan 1550 system from TOPTICA Photonics generates continuous terahertz waves using the method of near-infrared heterodyne conversion in which a photoconductive switch, integrated with an antenna structure, is illuminated by the combined output of two near-infrared lasers. By applying a bias voltage across the photomixer, a metal-semiconductor-metal structure, terahertz waves are emitted by the antenna exactly at the difference of frequency of two lasers. The detector is a second photomixer which is illuminated concurrently by the combined output of same two lasers as well as by the terahertz wave propagating from the transmitting antenna. The resulting photocurrent, $I_\mathrm{ph}$ depends on both the amplitude of the terahertz electric field, $E_\mathrm{THz}$, and the phase difference, $\Delta\varphi$, between the laser beat signal and terahertz radiation \cite{Toptica, stanze2011compact, deninger20152}. 

\section{Measurement Setup}
The bistatic measurements setup is assembled on a simple optical breadboard resting on a optical bench. The transmitter and receiver are mounted on circular arms made of aluminium. The transmitter (Tx) arm is fixed with the optical table, whereas the receiver (Rx) arm is bolted to a \SIadj{350}{\milli\meter} aperture rotational (R1) stage (Standa:8MRB450-350-60). The sample holder is fixed on an another smaller rotational (R2) stage (Standa:8MR174-11-20) to rotate the sample around its own axis. The R2 stage is mounted on a translational XY stage (T1) with \SI{25}{\milli\meter} of throw. The center of the parabolic mirror located inside the Tx and Rx housings is located at a distance of \SI{77.5}{\milli\meter} from the sample center plane. These \ang{90} off-axis parabolic mirrors produce a collimated beam with an \SIadj{25.4}{\milli\meter} aperture. Given that aperture size, the far-field distance is approximately \SI{43}{\centi\meter} at \SI{100}{\giga\hertz}. These measurements are therefore made in the transition region between the near- and far-field. We have not conducted measurements at significantly larger distances. Although this system supports both transmission and reflection measurements, we only describe results from reflection measurements in this paper.

\begin{figure}[t!]
    \centering
    \includegraphics[width=0.4\linewidth]{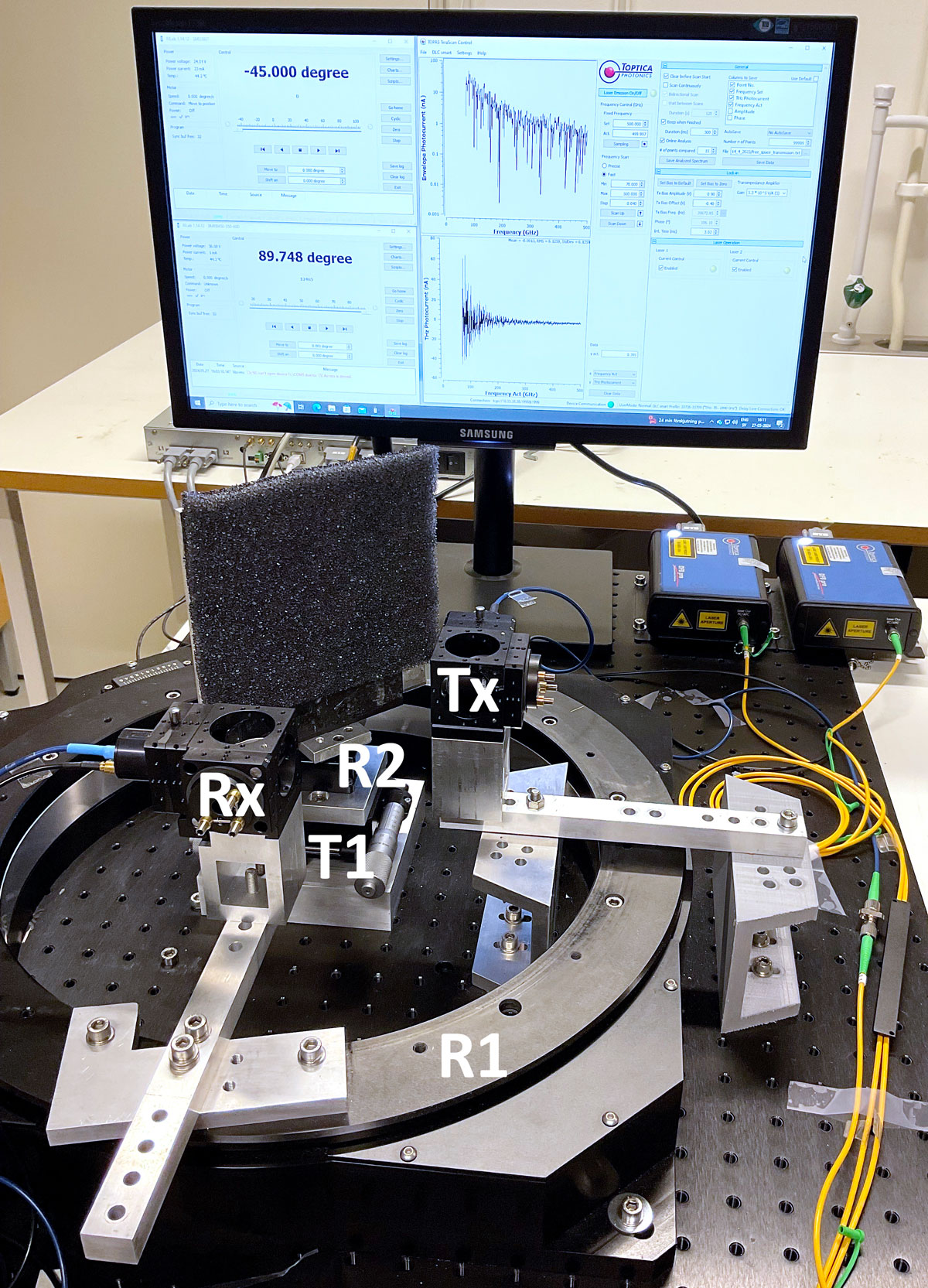}
    \caption{The Terascan 1550 system mounted in a custom reflectometry system composed of two rotation stages and a small linear translation system. The diagram shown in Figure~\ref{fig:diagram} represents a top-down view of this system.}
    \label{fig:measure_setup}
\end{figure}

The sample is mounted on a custom sample holder so that a rotation of R2 will vary the angle of incidence. Rotation of the receiver arm, which is mounted on the R1 stage, then changes the receiving angle and let us evaluate both specular and non-specular reflections. Due to the finite size of the transmitter and receiver assemblies, however, it is impossible to measure specular reflection for angles of incidence lower than about \ang{15}. Measurements of S- and P-polarization reflection response can be made by rotating the photomixer modules in the Tx and Rx housings. Note that P-polarization has its electric field oscillating parallel to the plane of incidence while S-polarization is perpendicular to this plane. This also implies that the electric field of an S-polarized wave is oscillating parallel to the sample surface plane as the signal from the Tx module hits the sample. A diagram of the measurement setup is shown in Figure~\ref{fig:diagram}. A photograph of the setup with a sample of an HR10 absorber is shown in Figure~\ref{fig:measure_setup}. The photograph omits an absorber that is strategically placed between the Tx and Rx module to minimize spillover that would otherwise impact the dynamic range of our measurements.

\subsection{System Characterization}

In this work, we present measurements covering 100 to \SI{500}{\giga\hertz} sampled at \SIadj{40}{\mega\hertz} intervals. Each frequency sample is based on \SI{3}{\milli\second} of integration time such that each scan through the full frequency range takes approximately \SI{320}{\second}. Measurements with a mirror surface in the sample location allows us to evaluate the dynamic range of our measurements as a function of integration time. We find that for \SI{3}{\milli\second} of integration time our dynamic range is below $-55$ and \SI{-70}{\decibel} at the low end of high end of our frequency measurement range, respectively. The dynamic range is calculated according to $20\log (I_\mathrm{noise}/I_\mathrm{signal})$ where $I_\mathrm{noise}$ is estimated by sampling the Rx signal, when no sample is mounted on the sample holder and $I_\mathrm{signal}$ is estimated with a flat mirror. The dynamic range estimates are consistent with statements made in the TOPTICA user manual.\cite{Toptica} %To ensure that the fidelity of the measurements is not compromised by the shorter integration times, we compared the response of the HR10 absorber for different integration time as shown in Figure~\ref{alignment}. 

To ensure accurate and reliable results, the following alignments are carefully maintained throughout our measurement process: (1) The transmitter and receiver antennas are at the same elevation from the breadboard; (2) the rotation axis of the R2 stage should be at the geometrical centre of the transmitter and receiver arm; (3) the rotation axis of the R2 stage lies in the plane of the sample surface; and (4) for zero angle of incidence, the axis perpendicular to the sample plane and the axis of incident waves from the transmitting antenna are co-linear. For reference, an aluminum plate is used to align the system in accordance with these conditions before quantifying absorber samples. After the alignment process, the sample, backed by an aluminum plate, is placed in the sample holder while the T1 stage facilitates adjustment for sample thickness. To mitigate unwanted reflections, the test sample is surrounded by an AN72 absorber. To check alignment, initially specular reflection response for an aluminum mirror was measured and irrespective of angle of incidence, similar response (in terms of photocurrent) was measured.

\section{Reflection Measurement}
\label{sec:reflection_measurement}

For reflection measurements the angle of incidence, $\phi_1$, was varied from \SI{15}{\degree} to \SI{45}{\degree} with a step size of \SI{7.5}{\degree}. For specular measurements, the angle of the receiver module, $\phi_2$, is configured such that $\phi_1 = \phi_2$ (see Figure~\ref{fig:diagram}). To probe non-specular reflections, we scan through a range of angles, $\phi_2 = 15, 22.5, 30, 37.5, \SI{45}{\degree}$, by moving the receiver module for a fixed angle of incidence. These five receiver angle configurations, $\phi_2$, are repeated for the different incidence angles, $\phi_1$ (see Figure~\ref{fig:diagram}). Note that both $\phi_1$ and $\phi_2$ are defined as positive angles relative to the sample normal line with $\phi_1$ and $\phi_2$ referencing the angle that the Tx and Rx modules make relative to the normal, respectively. 

The absorbers considered for this study are Tessellating TeraHertz RAM (TKRAM)\cite{TKRAM}, carbon- and iron-loaded 2850FT Stycast (from now on simply referred to as Stycast), in-house fabricated 3D printed pyramidal absorber (PMA) made from carbon reinforced polyamide and commercial absorber foams: AN72 \cite{AN72} and HR10\cite{HR10}. AN72 and HR10 are flat samples made up of polyurethane foam material impregnated with carbon black having thickness of 6 and \SI{10}{\milli\meter}, respectively. 
For the Stycast, we add carbon and iron powders with a mass ratio of 5\% relative to the Stycast. We add a 23LV catalyst to the mixture in an 8\% mass ratio relative to the combined dry mixture, before spreading it evenly across an aluminum plate. The flat epoxy sample, with an average thickness of \SI{6.7}{\milli\meter}, was allowed to cure at room temperature.

The TKRAM is made of carbon-loaded polypropylene plastic.\cite{TKRAM} A unit cell has a total thickness of \SI{15}{\milli\meter}, which consist of \SIadj{9}{\milli\meter} thick bulk material at the bottom that supports pyramids with square bases, each having a side length of \SIadj{4}{\milli\meter} and a height of \SI{6}{\milli\meter}. The PMA has been fabricated using an FDM-based 3D printer. It has an overall thickness of \SI{16}{\milli\meter}, which consists of an \SIadj{9}{\milli\meter} thick base and a pyramidal structure with a \SIadj{6}{\milli\meter} side lenght and a height of \SI{7}{\milli\meter}.   %As evident from the size of the pyramids, they serve a different purpose for each absorber, where for MMA pyramidal shape behave as a anti-refection coating and TKRAM pyramidal structure helps in increasing the interaction of incoming wave with the material by increasing multiple reflection.

The Terascan 1550 system measures the response in the form of photocurrent, $I_\mathrm{ph} \propto E_\mathrm{THz}\cos(\Delta\phi)$, where $E_\mathrm{THz}$ is terahertz electric field amplitude and $\Delta\phi$ is the phase difference between the terahertz wave and the optical beat.\cite{Toptica} To estimate the reflected power from this dataset, the photocurrent envelope of the measured response is determined through the identification of local maxima and their corresponding frequency, including midpoints. Using linear interpolation, the envelope over a specified range (100 to \SI{500}{\giga\hertz}) can then be calculated. The resulting envelope is considered as the reflection from the sample under test. For specular as well as non-specular measurements, the reflection response of the sample is normalised by the reflection from an aluminum mirror. The reflectance, i.e., the fractional power reflected into angle $\phi$ is found according to: 
\begin{equation}
    R(\nu, \phi) = \left(\frac{I_{\mathrm{S}}(\nu, \phi)}{I_{\mathrm{A}}(\nu, \phi_1)}\right)^2,
    \label{eq:specular}
\end{equation}
where $I_\mathrm{S}(\nu, \phi)$ is the measured photocurrent from the sample as a function frequency and receiving angle, while $I_\mathrm{A}(\nu, \phi_1)$ is the measured photocurrent found using an aluminum plate when $\phi_1 = \phi_2$ (see Figure~\ref{fig:diagram}). %$I_{\mathrm{S}$ is the reflection response from a sample at $i$ angle of incidence and $\theta$ angle of receiving. 
\section{Result and Discussion}
  
\subsection{Specular Measurements}

\begin{figure}
    \centering
    \includegraphics[width=0.9\linewidth]{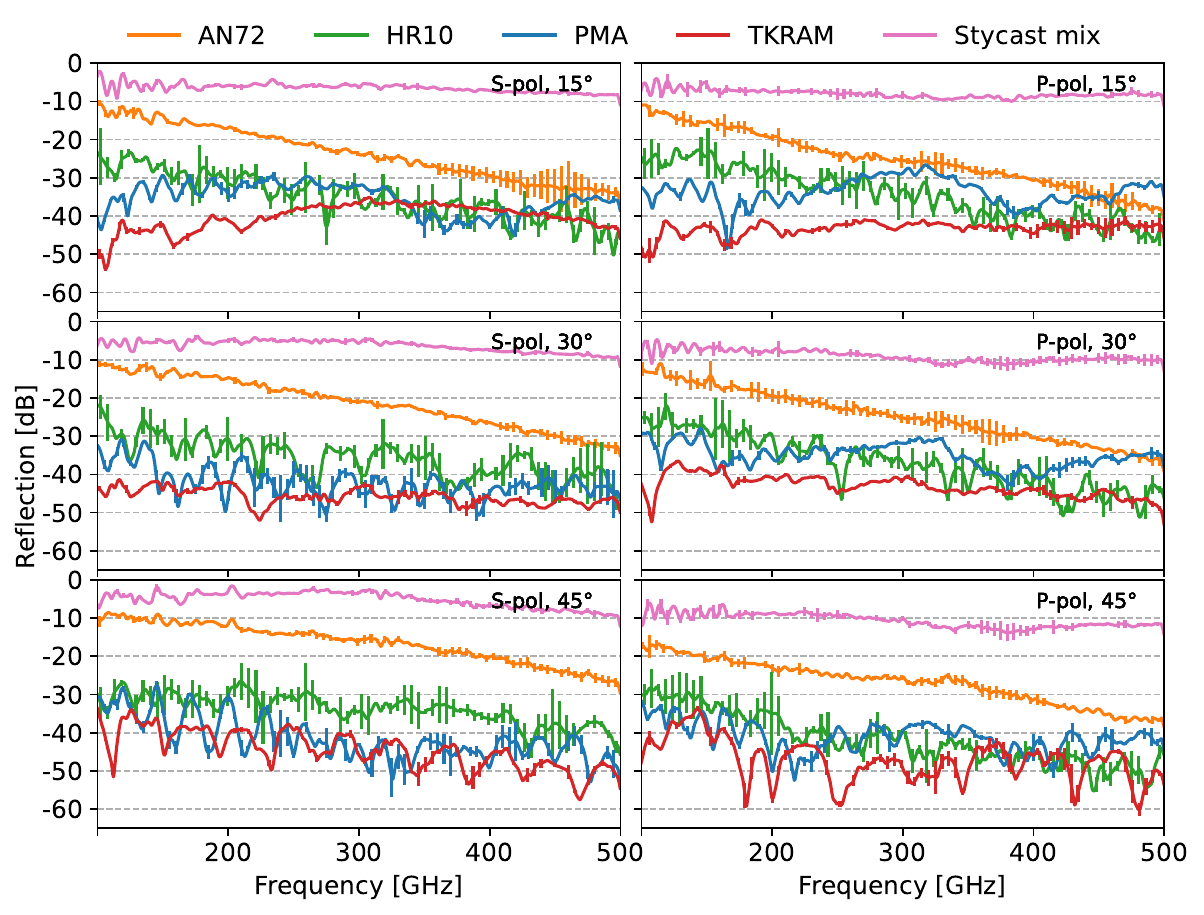}
    \caption{Comparison of specular reflection response from various absorbers at different angle of incidence ($\phi_1 = \ang{15}$, \ang{30}, and \ang{45}) for the two polarizations, S-polarization (left column) and P-polarization (right column).} 
    %Subplots (a), (c) and (e) shows the response $15^\circ$, $30^\circ$, $45^\circ$ incidence angle for S-Polarizarion, respectively. Similarly, subplots (b), (d) and (f) shows the same response for TM at the same angle of incidence }
    \label{fig:specular_comparison}
\end{figure}

Comparison of specular reflection measurements for the five absorber tiles covered in this work is presented for different angles of incidence and polarization in Figure~\ref{fig:specular_comparison}. Each sample is measured twice, with the sample rotated \ang{90} around the plane normal and reinstalled in the sample holder between each measurement. The curves represent the average response from those two measurements sampled at \SIadj{200}{\mega\hertz} intervals after having been smoothed with a filtering kernel that reduces fluctuations on scales smaller than \SI{1}{\giga\hertz}. The spread from the two measurements is shown in errorbars at \SIadj{5}{\giga\hertz} intervals. It is clear from this that some samples give a larger variation between measurements. We note that measurements are not noise-limited, but the errors represent a statistical statement about the repeatability of these measurements. 

TKRAM exhibits the strongest absorption overall across the different polarization angles and incidence angles. At near-normal incidence angles, the TKRAM specular reflectance does not exceed \SI{-37}{\decibel} in the 100 to \SI{500}{\giga\hertz} range for both polarizations and for a wide range of frequencies and incidence angles the specular reflectance is below \SI{-40}{\decibel}. We note that the lowest reflectance is found for high incidence angles ($\phi_1 = \ang{45}$), this can probably be expected given the large unit cell size of the TKRAM and the height of the pyramids. These results are in agreement with the study carried out by Lonnqvist et al.\ (2006)\cite{lonnqvist2006monostatic}. The results shows that for these TKRAM absorbers, specular reflection falls with an increase in angle of incidence. 

Overall, the 3D printed absorber, PMA, performs slightly worse than the TKRAM tiles. However, the reflected power never exceeds \SI{-29}{\decibel} for all frequencies and incidence angles. Significant variations in the reflected power as a function of frequency can be observed. This is related to the overall geometry of the PMA, both unit cell size and pyramid height. 

AN72, HR10, and doped Stycast resins are commonly used in mm-wave astrophysics. However the performance of these absorbers is seen to be quite different. A flat piece of carbon- and iron-doped Stycast has the highest reflection levels for all the angles and polarization's among all the samples under test, as expected. The lowest recorded reflection for Stycast is around \SI{-12}{\decibel} for both polarizations. Whereas, HR10 consistently shows lower than \SI{-25}{\decibel} reflection response in all the measurements conducted for this study. The AN72 absorber falls somewhere in between. We note that the foam-like absorbers, AN72 and HR10, show significant variation in the measured reflectance between measurements unlike the rest of the absorbers. This is especially true for the HR10. 
 
\subsection{Non-specular measurements}
There is no reason to expect that all reflected power from the transmitter will be specularly reflected. To gauge how much power is reflected across a broad range of incidence angles, we repeat our reflection measurements for five reflection angles (see discussion in Section~\ref{sec:reflection_measurement}).

\begin{figure}
    \centering
    \includegraphics[width=0.9\linewidth]{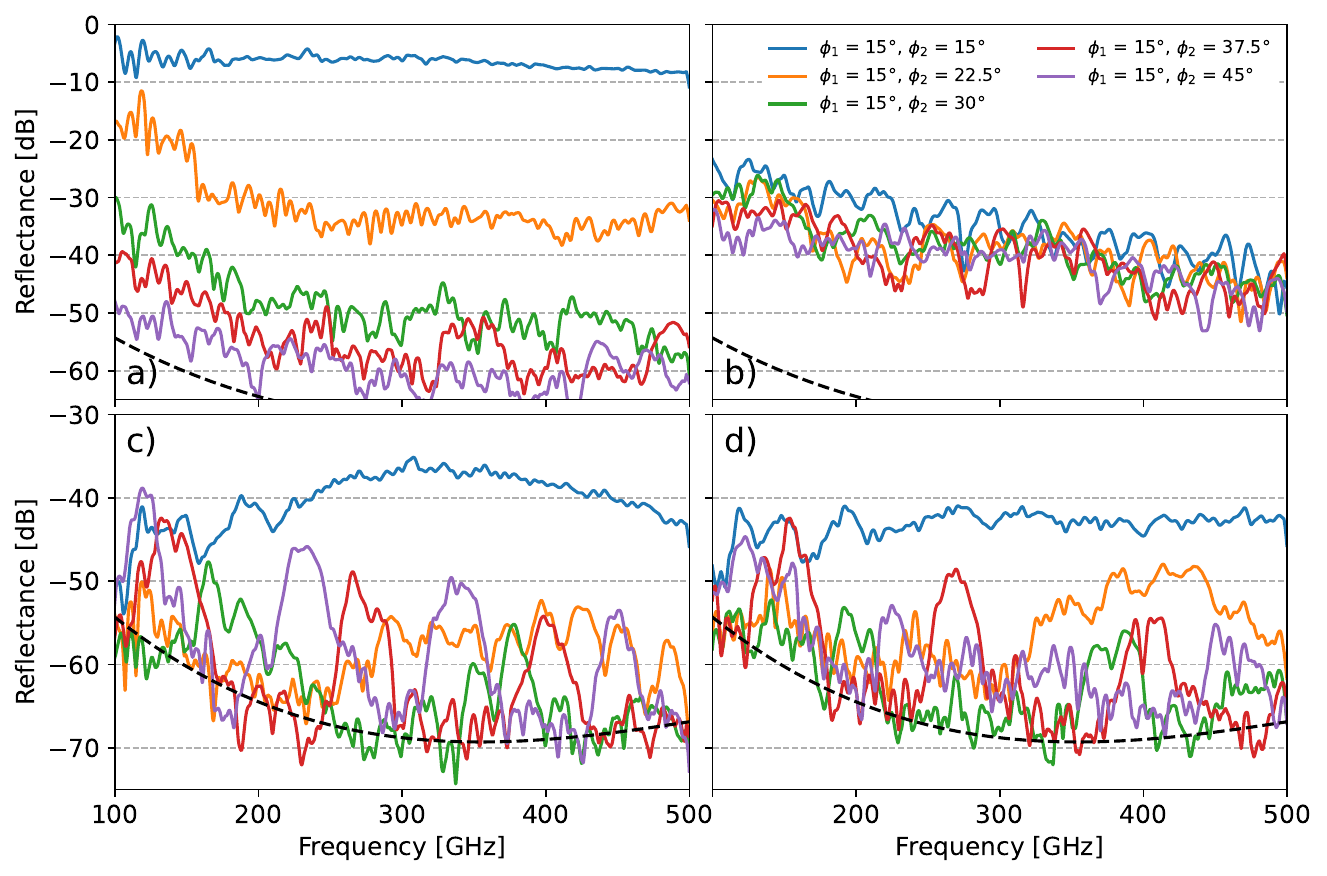}%{fig/scattering_comparison_jegv1.pdf}
    \caption{Each of the four panels shows the reflected power for an incidence angle of $\phi_1 = \ang{15}$ and five values of $\phi_2$ ranging from 15 to \ang{45} (see Figure \ref{fig:diagram}). a) The reflected power from Stycast in S-polarization. b) The reflected power from HR10 in S-polarization. c) The reflected power from TKRAM in S-polarization. d) The reflected power from TKRAM in P-polarization. Each value of $\phi_2$ is represented by a different colored curve with specular reflection in blue. The black dashed line represents our estimate for the measurement noise floor as a function of frequency given the \SIadj{3}{\milli\second} integration time.}
    \label{Scattered_comparison}
\end{figure}

Figure~\ref{Scattered_comparison} shows the reflected power as a function of frequency for three of the absorber candidates. The results are plotted for an incidence angle of $\phi_1 = \ang{15}$ with $\phi_2$ values ranging from 15 to \ang{45}. For the flat Stycast sample, we see that most of transmitted power is reflected specularly ($\phi_1 = \phi_2$), as expected, with less than \SI{-60}{\decibel} reflected power for large values of $\phi_2$ at the high frequency range. Measurements of HR10 however, suggest that the incident power is scattered rather uniformly into a range of angles. This is caused by the low density foam like structure of the HR10. For TKRAM, which has a unit cell size of \SI{4}{\milli\meter} square, we see variation between polarizations and a significant frequency dependence in the amount of power that is reflected non-specularly. This should be expected given the absorber geometry. 

\section{Conclusions and discussion}
The system described in this work is able to characterize the reflectance of absorbers across a wide range of frequencies, incidence angles, reflectance angles, and polarizations, with a dynamic range spanning more than 6 orders of magnitude. Combined with measurements of complex dielectric constant based on transmission measurements, this system helps to advance our understanding of broadband microwave absorber performance for future CMB experiments. 

Future work will provide a more detailed view of reflected power as a function of reflectance angle and temperature. Such measurements can be combined to provide an overview of the total reflected power from different absorbers used in current and future mm-wave optical systems. By combining such measurements with detailed simulations, we can work to improve our communities understanding of broadband absorbers at these frequencies.

%produce repeatable measurements with high dynamic range across the frequency range that is crucial for experiments studying the cosmic microwave background. 

%We show that the system is capable of 

%The system can be used to quickly scan through a wide range of angles to assess the integrated non-specular scattering from various absorber candidates. 

\acknowledgments % equivalent to \section*{ACKNOWLEDGMENTS}     
This work is supported by the Swedish National Space Agency (SNSA/Rymdstyrelsen). JEG acknowledges support from the Swedish Research Council (Reg.\ no.\ 2019-03959). JEG and TJLJG acknowledge support from the Icelandic Research Fund (Grant number: 2410656-051). Funded in part by the European Union (ERC, CMBeam, 101040169). Views and opinions expressed are however those of the author(s) only and do not necessarily reflect those of the European Union or the European Research Council Executive Agency. Neither the European Union nor the granting authority can be held responsible for them.

% References
\bibliography{report} % bibliography data in report.bib
\bibliographystyle{spiebib} % makes bibtex use spiebib.bst

\end{document}